\documentstyle[12pt]{article}
\def\beq{\begin{equation}}
\def\eeq{\end{equation}}
\def\rg{$R_{s}(p)$}
\def\fg{f_{s}(p)}
\def\za{z_{a}(s)}
\def\zh{z_{h}(s)}
\def\ns{n_{s}(p)}
\def\zd{z_{d}}

\begin{document}

\titlepage
\begin{flushright}
Oxford OUTP-93-22P\\
CERN-TH-6995/93\\
TUM-TH-159/93 \end{flushright}
%\vspace{0.7cm}

\begin{center} \bf

LARGE SCALE STRUCTURE FROM BIASED NON-EQUILIBRIUM PHASE
TRANSITIONS\\  -- PERCOLATION THEORY PICTURE\\
\rm
\vspace{3ex}
Z. Lalak${}^{a,b}$, S. Lola${}^{a}$, B. A. Ovrut${}^{c}$
 and G. G. Ross${}^{a,}{}\footnote[1]{\rm SERC Senior Fellow}$\\
\vspace{3ex}
${}^{a}$ Dept. of Physics \\Theoretical Physics \\ University of
Oxford \\
 1 Keble Road, Oxford, OX1 3NP\\
\vspace{2ex}
${}^{b}$ Physics Department\\ Technische Universit\"{a}t
M\"{u}nchen\\ D-85748 Garching\\
\vspace{2ex}
${}^{c}$ Theory Division CERN\\
CH-1211 Geneva 23\\
\vspace{2ex}

ABSTRACT
\end{center}
\noindent{
We give an analytical description of the spatial distribution of
domain walls produced during a biased nonequilibrium
phase transition in the vacuum state of a light scalar field. We
discuss in detail the spectrum of the associated cosmological energy
density perturbations. It is shown that the contribution coming
from domain walls can enhance the standard cold dark matter
spectrum in such a way as to
account for the whole range of IRAS  data and for the COBE
measurement of the microwave background anisotropy. We also
demonstrate that in case of a biased phase transition which
allows a percolative description, the number of large size domain
walls is strongly suppressed. This offers a way of
avoiding excessive  microwave background distorsions due to
 the  gravitational field of domain walls
present after decoupling.}

\newpage

\noindent{\rm \large 1. Introduction}
\vspace{2ex}

Over the  years there has been considerable effort put into
explaining  the observed large scale structure of the Universe.
Recently,  interest in the subject  has
been greatly stimulated by new data coming from the positive
signal from  COBE [1] and from the extensive IRAS survey [2].
 Perhaps one of the most interesting
new conclusions is the lesson that the standard cold dark matter
(CDM) scenario with a Harrison-Zeldovich spectrum of primeval
fluctuations apparently cannot alone account
for the whole spectrum of observed energy density fluctuations
[3],[4]. If one normalizes the CDM spectra so as to explain the
data on very small scales,  then there are
not sufficient fluctuations predicted at larger scales to be
consistent with  the large scale IRAS results, let alone to offer
any possible explanation of the COBE
measurements. The discrepancy is  significant, despite the fact
that data on larger scales also have large errors,  of the order
of 20\% of the critical density at 40--50 Mpc and about 10\% of
the critical density   at 100 Mpc
(assuming $H_0 = 100\; {\rm km} {\rm sec}^{-1} {\rm Mpc}^{-1}$).
On the particle
theory side, there have been several attempts to supply more
power to the long  wavelength parts of the theoretical spectrum
[3],\cite{lr}. Some of them replace pure  CDM with a
mixture of CDM, hot and warm dark matter while others use
topological defects created
during cosmological phase transitions to depart from CDM-like
predictions. In this work, we perform a detailed analysis of
density perturbations  produced by biased, non-equilibrium phase
transitions
in the vacuum state of a light weakly interacting scalar field.
We consider domain walls formed
during such a transition, and argue that under quite general
circumstances  they act as bag-like seeds for density
inhomogeneities of the baryonic and  cold dark matter
filling the Universe. Our statistical method, based on
percolation theory,  allows us to formulate a quantitative
picture for structure formation, and to  perform an
analytical calculation of the theoretical spectrum. In a natural
way, we  obtain predictions
 for the overdensity arising from these domain walls. When this
is added to  the CDM spectrum,
 we obtain a remarkably accurate prediction for the total
density fluctuations which is consistent with the entire range
of the IRAS  data and, simultaneously, with the COBE
measurements.
We stress that these results are obtained using domain walls
which are perhaps the most natural kind of topological defects,
easily accomodated by a large  class of realistic models  [6].
The scalar field we introduce is assumed  to interact only
gravitationally with the rest of matter, and to have very weak
self interactions. Hence this field is never in thermodynamic
equilibrium. For this reason our theory escapes all previous
no-go results for cosmological domain walls. We will, henceforth,
refer to such a scalar field as a  ``structuron'', since its main
physical role is the creation of large scale  structure.
Furthermore, although the associated phase transition encompasses
the so-called late-time phase transitions, it is actually much
more general. In particular, our transition can occur at any
time, such as well before  photon decoupling, in distinction to
late-time transitions which only occur long after recombination.
A further difference is that the non-equilibrium nature of our
transition allows us to bias the preference of the theory  to
choose one vacuum state over another. In the original late-time
transitions the vacuum states were populated with equal
probabilities. For these reasons, we will refer to the out of
equilibrium phase transitions of the structuron as ``structure
transitions''. In this paper the structure transition will in
fact occur prior to photon decoupling.

The paper is organized as follows. In sections 2 and 3 we
recapitulate basic facts about domain walls and elements of
percolation theory relevant in cosmology. In section 4 we sketch
the generic inflationary mechanism for producing domain wall
patterns leading to realistic energy density spectra. In section
5 we calculate analytically these spectra while in section 6  we
perform fits to observational data in the context of realistic
particle  physics models. In section 7 we comment on the possible
sources of microwave  background distortions which arise in our
scenario. Finally, section 8  contains a brief summary and
presents our
conclusions.
\\
\\
\noindent{\rm \large 2. Domain Walls}
\vspace{2ex}

Domain walls, which are planar defects in the vacuum alignment
over space, can appear if the manifold ${\cal M}$ of degenerate
(or nearly degenerate)  vacua of the
theory is disconnected, that is, when $\pi_{0}({\cal M})$ is
nontrivial.  This is in
general the case if there is a discrete symmetry in the theory.
This symmetry can be fundamental as, for example, is the $\phi
\; \rightarrow \; -\phi$ discrete symmetry in the double-well
potential \beq
V(\phi)= \frac{\lambda}{4} \upsilon^4 (\frac{\phi^2}{\upsilon^2}
-1)^2 \eeq
It may also arise dynamically during the breakdown induced by
instanton effects of a
continuous  symmetry, such as $\phi \; \rightarrow \; \phi + c$,
for example by QCD-like gauge
instantons or Dirac-type monopoles in antisymmetric tensors.
Moreover, discrete symmetries are often  left over after an
explicit breaking of a continuous symmetry due to Yukawa-type
interactions as in the case of classical pseudogoldstone bosons
or due to higher-order nonrenormalizable interactions which often
happens in string-inspired models\cite{lr}. In all these cases
the resulting multi-well potential is of the form
\beq
V = V_{0} [ \cos \left(\frac{ \phi}{\upsilon}\right) +1 ]
\label{e1}
\eeq
which breaks the continuous symmetry down to the discrete
symmetry $\phi \; \rightarrow \; \phi + 2 \pi m \upsilon, \;
m=1,2, \ldots$. The potential (\ref{e1}) has a finite number of
distinct degenerate minima if $\phi$ is an angular-type degree
of  freedom or an infinite number of inequivalent minima if the
manifold spanned  by $\phi$ is noncompact. The results of this
paper are generic to all theories of the above types. In fact,
it is not even necessary that the  nonequivalent
minima be
exactly degenerate, so the symmetries we consider may be only
approximate  ones, although non-degeneracy may change the details
of the wall evolution\cite{lr}.
However, as we have noted, many reasonable models give rise
naturally to cosine potentials. Therefore we will use
potential (\ref{e1})
in the explicit calculations in this paper.  As  will become
clear below,  even though such potentials can have a large number
of inequivalent  degenerate
minima, in this paper we need only consider two adjacent vacua.

The equation of motion corresponding to (\ref{e1}) admits kink-
like, static domain wall solutions that interpolate between two
adjacent vacua. Locally, we can consider these walls as lying in
the $x-y$ plane. In this case the static domain wall solution
between the two vacua  $\phi_{+,-} = +\pi \upsilon, \; -\pi
\upsilon \;$ is given by
\beq
\phi_{wall}(z;z_0)= 2 \upsilon
\, \arctan ( \sinh ( \frac{z-z_0}{\Delta}))
\label{eq:e15}
\end{equation}
where $z_0$ is the arbitrary location of the wall,
$\Delta$ is its width,
\begin{equation}
\Delta = \frac{\upsilon}{ \sqrt{V_{0}}} =
m^{-1}
\label{eq:e16}
\end{equation}
and $m$ is the mass of $\phi$ evaluated at any minimum.
The surface energy density of
the wall is easily computed to be
\begin{equation}
\sigma = \int_{-\infty}^{+\infty}
V_{0} \left[\cos \left( \frac{ \phi_{wall}}{\upsilon}\right) +1
\right] \, dz  = 8 \, \upsilon^2 m
\label{eq:e17}
\end{equation}
Of course, an arbitrary spatial superposition of domain walls,
such as  that produced by the physical mechanism described in
this paper, is not a solution to the equation of motion and
cannot be stable. However, we argue that such a superposition
represents physically  meaningful initial conditions, whose
subsequent evolution is governed  by the   dynamics of the
theory. Subject to this dynamics, the initially static domain
walls  acquire non-zero velocities  and oscillate under their
surface tension.  The basic
features of domain wall evolution relevant for our calculations
are discussed
and applied in sections 5 and 6.
\\
\\
\noindent{\rm \large 3. Basics of Percolation Theory}
\vspace{2ex}

Consider three-dimensional
space partitioned
into cubic lattice sites with
lattice spacing $\Lambda$. We will assume that at each lattice
site  the physical system can be in one of two vacua, which
we will call $(+)$ and $(-)$ respectively.
The probability that
a lattice site is in the $(+)$ vacuum
is denoted by $p$ where  $0 \leq p \leq 1$,
while the probability that a lattice site
is in the $(-)$ vacuum is $q=1-p$.
If there is  no correlation  between the
vacuum structures at  any two different lattice
sites, it is possible to calculate
the spatial distribution of the two vacua, and, hence, the
spatial  distribution of
domain walls by applying three-dimensional
percolation theory.

In [7] it has been shown that as long as
$p < p_{c}$, where $p_{c}=.311$ is the
critical probability for a cubic lattice in three dimensions,
the $(-)$ vacua lie predominantly
in a large percolating cluster (since necessarily $q>p_c$) while
the $(+)$
vacua are in finite
s-clusters. Here $s$ denotes the
number of nearest neighbour lattice sites that
are occupied by $(+)$. Two sites are considered nearest
neighbours if they  have a common link.
Also, as $p \rightarrow p_c$ from below, large pre-incipient
percolating clusters of $(+)$ vacua may begin to form in a finite
lattice. However, this happens only for $p$ very close to $p_c$.
Since, in our
discussion, we always take $p$ much smaller than $p_c$,
we need not discuss this subtlety further.

Let $\ns$ be the probability per lattice site that a given
lattice site is an element of an s-cluster. This is
a fundamental quantity, given by the ratio of the
total number of s-clusters, $N_{s}$, over the total number of
lattice sites, $N$.
An analytical expression for this quantity has been found, [7],
using scaling arguments and Monte Carlo simulations. The result
is \begin{equation}
\ns = .0501 s^{-\tau}\;e^{-.6299 (
\frac{p-p_{c}}{p_{c}}) s^{\sigma} [(\frac{p-p_{c}}{p})
s^{\sigma}+ 1.6679]}
\label{ns}
\end{equation}
where $\tau = 2.17$ and $\sigma=.48$.
The average radius of gyration for an s-cluster $R_{s}(p)$,  for
$p < p_{c}$ and $s > s_{\xi}$, is found to be
\begin{equation}
R_{s}(p) \equiv f_{s}(p) \Lambda =
 .702 (p_{c}-p)^{.322} s^{.55} \Lambda
\label{Rs}
\end{equation}
where
\begin{equation}
s_{\xi} = \left(\frac{.311}{|p-.311|}\right)^{2.08}.
\label{sz}
\end{equation}
It can also be shown, for $p<p_c$ and $s
\stackrel{\textstyle >}{\sim}  5$, that every s-cluster
 has a boundary
composed of
\beq
t_{s} = \left(\frac{1-p}{p}\right)s
\label{ts}
\eeq
$(-)$ vacuum sites.

Finally, we want to make one important remark. One can easily
check using  the formulae of this section that, on a given
lattice,
 the number of s-clusters
falls rather quickly with growing s. Hence there is an ${\rm
s}_{max}$ such that the total number of ${\rm s}_{max}$-clusters
is 1. In other words,  formation of clusters with s much larger
than ${\rm s}_{max}$ is extremely  improbable. This means that
on a given lattice there exists an upper  statistical
cut-off on the size of observable clusters. This is, as discussed
in section  7, a very important property of our scenario which
allows us to avoid  excessive microwave background distorsions
due to large domain walls.  \\
\\
\noindent{\rm \large 4. Biased Non-Equilibrium Phase Transitions
in the Postinflationary Universe}
\vspace{2ex}

We want to consider a universe that for some reason, perhaps due
to the existence of an inflaton field, goes through a period of
inflation and then settles down into the standard Friedman-
Robertson-Walker (FRW) geometry. However, in addition, we
postulate the existence of another scalar,  $\phi$, not the
inflaton, which has a potential with several degenerate  (or
almost degenerate) minima. As we said previously, in this paper
we will assume that this potential is of the form (\ref{e1}).
Furthermore, we will assume  that
\beq
\frac{V_0}{\upsilon^2} << H_i^2
\eeq
and
\beq
V_0 << \upsilon^4
\eeq
where $H_i$ is the Hubble constant during inflation. The first
condition  implies that during inflation, and long after, the
potential force is much smaller than the cosmological friction, which
justifies neglecting the potential  until the universe is deep
into the FRW expanding phase. The second condition guarantees
that the $\phi$ field self couplings are weak. Since the only
other interactions $\phi$ experiences are gravitational, this
implies that $\phi$  is never in, or even near, thermal
equilibrium. How small $V_0$ or how  large $\upsilon$ is will be
determined by our fit to the $\delta \rho / \rho$ data,
and will be discussed below. As discussed above, we will call a
field with these properties
a ``structuron'', since the major role it plays in our scenario
consists solely  in inducing large scale structure formation.
Detailed analysis of fluctuation-driven
out of
thermal
equilibrium phase transitions in the post-inflationary universe  will
be presented in a separate paper [8].  Here we
recall the main ingredients of this physics that we will need in
the  discussion below. We begin by stressing, once again, that
the structuron  is assumed
to interact so weakly with itself and with other fields that it
never  comes close to thermal equilibrium.
An out-of-equilibrium scalar field $\phi$ living on an inflating
de Sitter space, observed over a physical volume $\ell^3$,
breaks into two pieces
\begin{equation}
\phi = \phi_c + \phi_q
\label{1}
\end{equation}
where $\phi_c$ satisfies the classical equation of motion
\begin{equation}
\ddot{\phi}_c + 3H \dot \phi_c + \frac {\partial V}{\partial
\phi_c} = 0
\label{2}
\end{equation}
(we have assumed that initial spatial gradient terms are
exponentially  driven to
zero by
inflation) with $V$ the potential energy and $H$ the Hubble
parameter, and $\phi_q$ represents de Sitter space quantum
fluctuations. During the inflationary  de Sitter period $H=H_i$.
However, equation (\ref{2}) is also valid after
inflation has ended and the universe is in the FRW expanding
phase.  During inflation, and long afterward, $H^2$ is very large
compared to the
curvature of the potential.  Hence, the $\frac{\partial V}{\partial
\phi_c}$ term
can be dropped
to lowest order and $\phi_c = \vartheta$ where $\vartheta$  is
an arbitrary
constant.  To
next order, there is a tiny damped  velocity
$\dot \phi_c
\sim \frac{V}{H \upsilon}$. Hence, during inflation, and long
afterward, $\phi_c$ is
to  very good accuracy an arbitrary constant.
As already mentioned, $\phi_q$
represents
quantum
fluctuations  of the scalar field in deSitter space.
These fluctuations
result in the formation of a weakly inhomogeneous
quasi-classical  random field. After inflation ends, the FRW horizon,
$\ell_c=1/H$, grows and fluctuations with scales less than the
horizon are smoothed out. Thus $\ell_c$ acts as an  UV cut-off in
the momentum distribution of this random field. In a spatial
region of length $l$,  the distribution
of the fluctuations around $\vartheta$ can be calculated and is
given by \begin{equation}
P(\phi)=\frac{1}{\sqrt{2 \pi} \sigma_{\ell}} \exp{(-\frac{(\phi -
\vartheta)^2}
{2 \sigma^2_{\ell}})}
\label{prob}
\end{equation}
where
\begin{equation}
\sigma^2_{\ell} = \frac{H_i^2}{4\pi^2} \ln \left(
\frac{\ell}{\ell_c} \right)
\label{3}
\end{equation}
Now, one can easily demonstrate [8],[9] that the longwavelength
components in the
Fourier
decomposition of the random field $\phi_q$ are important.
Therefore, the values of $\phi_q$
at any two distant points in the
region of interest are, in general, not independent. This means
 the de Sitter space fluctuations introduce correlations among the
values of  the random field
up to the maximal distance $H_{i}^{-1} e^{n}$, where $n$ is the
total  number of e-folds during inflation. The above conclusions
are strictly valid  for a field which has been present in the
universe since the beginning of  the inflationary epoch. The
situation with respect to long range correlations  changes if the
field $\phi$ is produced
 during inflation\footnote{This may occur in a number of ways.
For example,   a complex scalar field with a large initial
positive mass may, during inflation, undergo a phase
transition (not to be confused with the  structure transition)
to a spontaneously broken theory with a massive radial  field and
a pseudo Goldstone boson $\phi$.},
let us say at $x$ e-folds before the end of the inflationary
period.  Then, of course, the correlations in the random field
are produced only up to the scale ${\ell}_{x} =
H_{i}^{-1} e^{x}$, which may be much smaller  than $H_{i}^{-1}
e^{n}$.  If the inflationary period prior to the appearence of
the field $\phi$ is sufficently long, and provided that the
microphysics
 producing $\phi$ fulfills certain, rather nonrestrictive,
conditions [8], then  the decomposition (\ref{1}) remains valid.
In this case, however,  correlations in the random field are
restricted to  scales
smaller than $\ell_x$. After inflation ends, the scale $\ell_x$
expands  according to the expansion law of the subsequent FRW
epoch, and at  a given redshift $z \geq \zd$ equals approximately
\beq
\ell_{x} (z) = H_{i}^{-1}\;e^{x} \frac{T_r}{T_0}
\frac{1}{(1+\zd)^{1/4}} (\frac{H_0}{H(z)})^{1/2}
\label{lix}
\eeq
where $H_0$, $T_r$, $T_0$, and $\zd$ are the present Hubble
constant, the  temperature of reheating, the present microwave
background temperature  and the redshift at the beginning of
matter dominance respectively. Hence the distribution describing
the quasiclassical random  field $\phi_q$ at some redshift $z$
is that of (\ref{prob}) with  arbitrary $\vartheta$  and
$\sigma_{\ell}$ given by (\ref{3}) for  $\ell < \ell_{x}(z)$.
Note that as long as $\sigma_{\ell_x} \leq \pi \upsilon$, that
is
\beq
H_i \leq \frac{2 \pi^2}{\sqrt{\ln ( \ell_x / \ell_c ) }} \upsilon
\label{hi}
\eeq
which we henceforth assume, then the distribution (\ref{prob})
will only be significant over  the range of the two nearest
neighbour  vacua to $\phi_c =\vartheta$. Therefore, we need only
consider physics near these  two vacua, which we will call $(+)$
and $(-)$ respectively.

As long as $H$ is much larger than $V_0^{1/2} / \upsilon$, the
potential  term in (13) can be ignored and the vacuum is well
defined by the distribution  (\ref{prob}) of fluctuations around
an arbitrary constant $\vartheta$. Eventually, long after
inflation, $H(t)$ decreases so
significantly that the potential term in the equation of motion
becomes equal to, and then begins to dominate over, the damping
term.
We denote by $z_t$ the redshift at which the potential term
becomes non-ignorable. This is the redshift at which the phase
transition is triggered. We will always assume that $x$ is such
that $\ell_c (z_t) < \ell_x (z_t)$.

We are now able to derive the percolation theory picture for
domain wall formation. At $z_t$ the field configuration $\phi$
within an horizon size volume, $\ell_c (z_t)^3$, is uniform due to the
smoothing effects of the spatial derivative terms governing the
evolution of the field. There are, however, fluctuations
due to the components of the quasi-classical random field whose
wavelengths $\lambda$ satisfy $\ell_c (z_t) \leq \lambda \leq
\ell_x (z_t)$. These fluctuations within a horizon volume are
described by distribution (14) where $\sigma_l$ given in (15) is
to be evaluated at $z_t$. The field at $z_t$ feels the
potential and
must decide into which of the two vacua it will  roll under the
potential force. It is clear from Figure 1 that
the probability
that the system will roll into the $(+)$ vacuum is
\beq
p=\int_{-\infty}^{0} P(\phi) d \phi
\eeq
which, depending on the value of $\vartheta$, can lie anywhere
in the range  $0 \leq p \leq 1$. Since $\vartheta$ is arbitrary,
clearly $p$ is arbitrary. It follows that the probability that
the system will roll into vacuum $(-)$  is $q=1-p$.

After the phase transition is triggered, $\ell_x (z)$ grows
according to the FRW expansion given in (16). The distance to
the horizon, $\ell_c (z)$, also grows, but its rate of expansion
is faster. Now, after $z_t$, the $\phi$ field rolls down the
potential, oscillates and then settles in either the $(+)$
vacuum or the $(-)$ vacuum. The redshift at which the system
settles into one of these vacua is denoted by $z_f$, and can be
considerably smaller than $z_t$. It is at $z_f$ that truly
topologically stable domain walls form. There are now two
possibilities. The first is that $\frac{\ell_x (z_f)}{\ell_c
(z_f)} \stackrel{\textstyle < }{\sim} 1$. In this case, the
causal horizon at $z_f$ has ``gobbled up'' the region of $\phi$
correlations that had surrounded it at $z_t$. In this case,
there are no correlations larger than the horizon at the time of
domain wall formation and percolation theory can be employed.
This is the case we will discuss in this paper. The second
possibility is that $\frac{\ell_x (z_f)}{\ell_c
(z_f)} >> 1$. In this case, domain walls with size larger than
$\ell_x (z_f)$ still have a distribution given by percolation
theory. However, at smaller scales there are non-zero
correlations between causal horizons at the time of domain wall
formation and, hence, percolation theory must be modified. We
will discuss this second possibility in a later work.

We may conclude that, long after inflation has
ended,  the vacuum state of the system goes through a phase
transition, choosing vacuum $(+)$ with arbitrary probability $p$
or  vacuum $(-)$ with probability $q=1-p$. We emphasize that due
to the  nonequilibrium nature of this transition $p$ can be
arbitrary and  need not, as in the case of thermal equilibrium,
be equal to 0.5.

It is important to determine the value of the Hubble parameter
at  the redshift, $z_t$, at which the phase transition is
triggered.  The redshift at which the $\phi$ moves toward a
definite vacuum is  fixed by the condition that at that redshift
the field should be able  to change its value by the amount
comparable to the distance between the two minima within one
Hubble time
\beq
\frac{1}{H(z_t)} \; | \dot{\phi} | \approx \pi \upsilon
\eeq
Also, it is not hard to show that at $z_t$ defined in this way
all the terms in the equation of motion are comparable.  This
ensures that at $z_t$
\beq
3 H(z_t) \dot{\phi} \approx -2 \frac{\partial V}{\partial \phi}
\eeq
where the factor of 2 on the right hand side takes care of the
$\ddot{\phi}$ term. Combining these two results yields
\beq
H(z_t)^2 \approx \frac{2}{3 \pi \upsilon} | \frac{\partial V}{
\partial \phi} |
\label{hzt}
\eeq
{}From what has been said about spatial correlations due to
inflation, it  follows that
 the fundamental length scale of the $(+)$ and $(-)$ vacuum
domains,  or the lattice spacing, is given by $H^{-1} (z_f)$.
Let us define a parameter $\alpha \equiv H (z_t) / H (z_f) =
\ell_c (z_f) / \ell_x (z_t)$. Also, since we prefer to
refer to $z_t$, we will somewhat perversely denote
$H^{-1} (z_f)$ by $\Lambda (z_t)$. It follows that
\beq
\Lambda(z_t) =\frac{\alpha}{ H(z_t)}
\eeq
where we will keep $\alpha $ as a free
parameter in the forthcoming formulae. We will  always refer to
$z_t$, the difference between $z_t$ and $z_f$ being  encoded in
the factor $\alpha$. If we take a specific value of $\alpha$,
then
using formula (\ref{lix}) and assuming, for example, that
$\ell_x (z_f) \approx \ell_c (z_f)$ one can estimate the time when the
phase  transition producing $\phi$ occured. It is not difficult
to see that, with good reheating after inflation, $\alpha =
O(10)$ requires $x=O(60)$.

Finally, we shall determine the size of our lattice. We are
interested in the portion of the Universe at $z_t$ which
coincides with the present day visible horizon ($L(z=0)=6000
h^{-1}$ Mpc). Hence we need only consider the finite lattice
whose linear size  is that of the present horizon scaled back to
$z_t$
\beq
L(z_t)=L(0)/(1+z_t)
\eeq
The total number  of  lattice sites $N=(L(z_t)/ \Lambda(z_t))^3$
is then easily found to be
\beq
N=(\frac{H(z_t)}{H_0} \frac{1}{\alpha (1+z_t)})^3
\label{nn}
\eeq
\\
\\
\noindent{\rm \large 5. Energy Density Perturbations}
\vspace{2ex}

Let us assume that the fundamental vacuum domains
 defined in
the previous section form a cubic lattice. It follows from the
discussion  in the previous section that, at each lattice site,
the system is in the $(+)$ vacuum with arbitrary probability $p$
where $0 \leq p \leq 1$, or in the $(-)$  vacuum with probability
$q=1-p$. Furthermore, the choice of vacuum state at any two
different lattice points is uncorrelated.
This situation
 defines a two-state percolation theory on our lattice.
Moreover, we assume that at every link which separates two
different vacua a domain wall forms with probability close to
unity, as explained in [7].  Having at our disposal all the tools
of percolation theory described in section 3,  we are in a
position to analyse the  complete spectrum
of density perturbations caused by these domain
walls over a wide range of length scales.

In the following, we provide a detailed analysis for the strongly
biased case where $p<p_c <q$. More specifically, we consider
$0.1<p<0.2$, since results obtained in this regime have
relatively simple structure and may be directly compared with
experimental data as given in [4].

It has already been explained in [7] that the spatial structure
of  the percolation
pattern in this interesting range of $p$ is rather
straightforward. There are  compact bubbles of less favourable
vacuum $(+)$ surrounded by a sea of vacuum $(-)$. Each bubble has
a (fractalized) surface composed of a domain   wall which stores
a significant amount of energy.
It is reasonable to assume that the walls assume their quasi-
static shape  sometime between $z_t$ and $z_f$. From that point
on, we can trace the  evolution
of domain wall ``bags'' and the energy deposited in those bags.
Consider  a bubble of $s$ nearest neighbour $(+)$ vacua,
 the s-cluster in terms of percolation
theory. Its mean radius is well characterized by the average
radius of gyration \rg . Since, at $z_t$, the causal horizon is
the lattice size,  this radius is initially larger then the FRW
horizon.
However, the horizon radius grows at a faster rate than that of
the bubble (whose radius just
grows linearly with the expansion). Therefore, at some
redshift $\za$, the bubble comes
within horizon. From that point on the dynamics of the bubble
becomes important. As indicated by Widrow [10], a subhorizon-size
bubble should first shrink under its surface tension, then
undergo a few cycles of oscillations, finally
radiating  its energy in the form of scalar waves.
 Still, as the analysis of the spherical collapse model
indicates,  these bubbles are around sufficiently long to induce
local inhomogeneities in the background medium. These
inhomogeneities have the form of overdense  regions
composed of baryonic and dark matter gravitationally attracted
to the bag  wall. The energy that is stored in the domain wall
surrounding an s-cluster at redshift z is easily computed from
(\ref{eq:e17}) and (\ref{ts}) to be
\beq
E_s = f t_s \sigma \Lambda(z)^2
\eeq
where f satisfies $1 \leq f \leq 6$. Since, in this paper, we are
concerned  with small $p$, we shall use a moderate value of
$f=3$. A discussion of the  appropriate value of f would take us
beyond the scope of this paper. Happily, the value of f does not
substantially alter our quantitative or qualitative conclusions.

The spherical collapse model allows us to estimate the total mass
accreted  onto a seed
of energy $E_s$ by the time it disappears.
The result is
\beq
M_s = \gamma_s E_s
\eeq
where
\beq
\gamma_s = (\frac{4}{3 \pi})^{2/3}
 (\frac{1+z_{app}}{1+z_{dis}})^{4/3}
\label{col}%5.5
\eeq
and $z_{app}$ and $z_{dis}$ are the redshifts at which the seed
appears and  disappears respectively.
Note that $z_{app}$ and $z_{dis}$ and, hence, $\gamma_s$ can
depend on s. Domain walls are fully formed at some redshift
$z_f$. As discussed above,  at this time all these walls are
outside the FRW horizon, and matter does not substantially acrete
to them. At redshift $\za$, however, the domain walls enter the
horizon and the process of the effective matter acretion begins.
Therefore, we will take $z_{app}= \za$.
The detailed analysis of the evolution of the quasi-spherical
domain wall  in the FRW universe with a large
cosmological damping term will be reported elsewhere [11]. In
general,   one expects that the
presence of cosmological friction due to the expansion should
slow down  the oscillations with respect to those observed in
flat space [10].  This gives a hint that the seed in the
form of the domain wall bag may be around sufficiently long to
cause the  collapse
of amounts of background matter significantly  larger than the
energy of  the original
seed. Looking at the equation of motion for a scalar field one
expects  that $\gamma_s$ becomes smaller for larger $s$, since
larger s-clusters  come within the horizon later, when the
friction term becomes smaller. On the other hand, however, the
larger $s$ clusters have larger initial radii which makes the
process of their ``collapse'' longer.
Numerical investigation of the behaviour of thick and thin
spherical
domain walls
in FRW background seems to confirm these expectations [11].
Hence, in this paper
 we
simply assume the  value of $\gamma_s$ to be independent of $s$
and of order 10, the order of magnitude  supported by numerical
analysis, although in  specific cases one can get much larger
values.

Putting all this together, some time after $\za$ but still at
large redshifts, we are given local clumps of background matter
corresponding to families of  s-clusters. We can easily compute
the mean energy density given by those clumps (we will continue
to call them s-clusters although the  original s-clusters are
already gone). To this end we have to know the mean distance
between s-clusters at the redshift of their formation.  In what
follows, we will take this redshift to be $\za$, the redshift
when the original s-cluster enters the horizon, in accord with the
discussion of the previous paragraph.
Using percolation theory, we know the number of s-clusters on our
lattice. It is given by $N_s = \ns N$, where $\ns$ and $N$ are
defined in (\ref{ns}) and (\ref{nn}) respectively.
Hence the mean distance between s-clusters at redshift $z_a(s)$
in physical units is
\beq
d_{s}(\za)=\Lambda (\za)
\frac{1}{{\ns}^{1/3}}=\frac{\alpha}{H(z_t)} \frac{
1+z_t}{1+\za} \; \frac{1}{{\ns}^{1/3}}
\label{dist}%5.6
\eeq
where $\Lambda(\za)$ denotes the lattice spacing at $\za$.  If
we scale the above
distance to the present time ($z=0$), then we will get the
present day  observable
scale over which the s-clusters contribute to the total energy
density, $\lambda (s)$.
Now, let us compute the energy density of s-clusters at $\za$.
This is given  by
\beq
\rho (s) |_{\za} = \gamma_s
 f \sigma \frac{1-p}{p} s \ns \Lambda^{-1} (\za)
\eeq%5.7
At this point, it is useful to compute $\za$ itself. It is easily
obtained  by equating the mean radius of gyration to the horizon
at that redshift. We find that
\beq
1+\za = \frac{1+z_t}{\alpha  \fg}
\label{zas}
\eeq%5.8
with
\beq
(1+z_t)^2 = \frac{H(z_t)}{ H_0} (1+\zd)^{1/2}
\label{ztt}
\eeq%5.9
where $z_d$ is the redshift when the radiation dominated epoch
ends and  matter domination begins.
The above formula holds for redshifts $\za$ larger that the
redshift of  the beginning of the matter dominance, but it will
shortly be clear that  that is the range of $\za$ of interest in
this paper.

We are interested in our Universe, which is flat on the average,
so we assume that the mean total energy density, which includes
energy of all the components of the Universe, walls, dark matter,
radiation, is equal to the critical  energy
density. One can easily express the critical energy density at
$\za$ in  terms of the present day critical density $\rho_0$
\beq
\rho_c (\za) = \rho_0 \frac{(1+\za)^4}{1+\zd}
\eeq%5.10
This formula is true only when $\za \geq z_{d}$, which we
henceforth assume to be the case.
Now we are ready to compute the energy density perturbation due
to the  s-clusters at $\za$. It is
\beq \left.
\frac{\delta \rho}{\rho} \right |_{\za}(s)= \gamma_s
\frac{f \sigma H(z_t)}{\rho_0 \alpha}
\frac{1-p}{p} s \ns \frac{1+\zd}{(1+\za)^3 (1+z_t)}
\eeq%5.11
{}From this point on, one should evolve the density perturbation
computed at $\za$ to the present time using the full Einstein
equations for  adiabatic (metric) perturbation, see for example
[12]. However, we can  estimate the present energy density
perturbation spectrum using well  known results in the following
approximate procedure.
For small values of probability p
the average distance between s-clusters
at $\za$, which is the scale of the perturbation produced by
those clusters, is larger than the horizon at $\za$. This average
distance  eventually comes  within the FRW
horizon at some smaller redshift which we call $\zh$. There are
two  possibilities to be considered. The first is that
$\zh$ happens to be larger than or equal to $\zd$. In this case,
$\zh$  is easily
computed to be
\beq
1+\zh= \ns^{1/3} (1+z_t) / \alpha
\eeq
and  the perturbation grows
between $\za$ and $\zh$ by a factor of
$\frac{(1+\za)^2}{(1+\zh)^2}$, then between $\zh$ and $\zd$ by
a logarithmic factor $1+2 \log ( \frac{1+\zh}{1+\zd})$, and
finally by the usual growth factor during the  matter dominated
epoch $1+\zd$. Putting everything together,  one gets in this
case
\beq \left.
\frac{\delta \rho}{\rho} \right |_{z=0} (s) = \gamma_s
\frac{\alpha^2 f \sigma}{\rho_0}
\frac{1-p}{p} s \ns^{1/3} \frac{H^{2}_{0}}{ H(z_t)}
 (1+\zd) \fg (1+2 \log \frac{
1+\zh}{1+\zd})
\label{lar}
\eeq%5.13
(note that the reduced Hubble constant $h$ does not appear
explicitly  either here or in (\ref{sm})).\\
The other possibility is that $\zh$ is smaller than $\zd$. In
this case  superhorizon growth of the perturbation in the
radiation dominated epoch  lasts till $\zd$ and then is followed,
as in the other case, by linear growth during matter dominance.
It follows that
\beq  \left.
\frac{\delta \rho}{\rho} \right |_{z=0} (s) = \gamma_s
\frac{ f \sigma}{\rho_0}
\frac{1-p}{p} s \ns H_{0} \fg \frac{1}{(1+\zd)^{1/2}}
\label{sm}%5.14
\eeq
As discussed above, the s which enters the formulae for
 overdensities has to be expressed in
terms of the corresponding physical scale $\lambda (s)$.
We find that
\beq
\lambda (s) = (1+\za )\Lambda (\za)
\frac{1}{{\ns}^{1/3}}=\frac{\alpha}  { H(z_t)}
  (1+z_t) \; \frac{1}{{\ns}^{1/3}}
\label{llam}
\eeq%5.15
The formulae (\ref{lar}) and (\ref{sm}) together with
(\ref{llam})  constitute the main result of this section.
They give the continuous spectrum of energy density fluctuations
due to the formation of domain walls during the biased
phase transition of the structuron field.
These fluctuations should be added to
those produced by the inflaton itself in order to obtain the
total  spectrum of fluctuations which may be compared with
observations. Numerical analysis of the spectrum and comparison
with QDOT [4] results are presented in the next section.
For the sake of completeness one should notice here that the
range of $\lambda$, or equivalently $s$, over which the analytical
results (\ref{lar}) and  (\ref{sm}) are  valid is in fact
restricted for two reasons. Firstly, because the validity of the
expression for $\ns$ which we are using here has been established
in reference [7] only over a finite, although wide, range of $s$.
Secondly, because the formula for \rg{} changes for  $s<s_{\xi}$
taking the form relevant for the critical region. These
limitations and assumptions about the range of $z$'s considered are
checked  a'posteriori for numerical examples discussed in the
following sections.       \\
\\
\noindent{\rm \large 6. Fits of wall-induced density fluctuations
to experimental data}
\vspace{2ex}

The extensive IRAS survey has allowed construction of detailed
maps and   quantitative analysis of the large scale structure of
the Universe out to roughly $140\;h^{-1}$Mpc. Using this data one can
construct the density field of galaxies and, by adding some
theoretical assumptions, one can infer the mass density
distribution and the pattern of peculiar velocities in  the
Universe [2].
These experimental results can serve as a gauge for theoretical
models  of large scale structure formation. In particular, one
can test the pure cold  dark matter model, [3],
 perhaps the most popular model over the last decade, and very well
motivated by fundamental
particle theory. In this paper we will use the analysis
presented in [4] based on the QDOT redshift survey [2].
This analysis shows that CDM with primeval
fluctuations of  Harrison-Zeldovich type, generated for example
during an inflationary  period in the early Universe,
cannot alone account for the whole  spectrum of energy density
perturbations. If one normalizes the CDM spectrum in such a way
that it accurately fits the observed spectrum at small scales, up to
$5\;h^{-1}$Mpc, then there is a statistically significant
disagreement at larger scales, where it is found that
the CDM prediction falls well
below the  observed spectrum. The plot in Figure 2 presents the
difference between  CDM simulations  and
observational data, as reported by Saunders et al. [4] (the
uncertainties associated with measurements at larger scales are
shown as two vertical lines).
The particular CDM spectrum which has been used is normalized in
such a way
that $b_I \sigma_8 = 0.69$ as found
for IRAS galaxies in [4]. Here the $b_I$ is the linear bias
factor (I stands for IRAS) and $\sigma_8$ is the rms matter
density  fluctuation
in spheres of radius $8\;h^{-1}$ Mpc, the quantities usually used
for  normalizing
theories of large scale structure. In this paper we choose to work
with   the inflationary CDM spectrum normalized so that it accurately
describes
small scale data instead of normalizing it, for
instance,  at the COBE point. We feel more comfortable with this
 normalization, as
the amount of data available in the range of smaller scales is extensive
with its structure carefully studied, and also because the
physical  length scales
 parametrizing the potentials discussed in this work correspond
to these smaller  cosmological scales.

The main numerical result of this paper is shown in Figure 3.
The curves (a),(b),(c) represent our fits of the domain wall-induced
overdensity spectrum (\ref{lar}), (\ref{sm})  in models with a potential in
the form (\ref{e1}), to the difference between pure CDM (with Harrison-
Zeldovich primeval fluctuations) and the observed rms energy
density spectra,  as given by QDOT
[2], [4],  on scales between 20
Mpc$\;h^{-1}$ and 80 Mpc$\;h^{-1}$  (for the simplicity of
presentation we put the reduced
Hubble constant h=1 from this point on)\footnote[2]{It has
already been noted  that the reduced Hubble constant $h$ drops out
from the formulae (\ref{lar}), (\ref{sm}). In fact it is present there only
implicitly through the value of $\zd$. On the other hand, it may
be seen from (\ref{llam}) that since $H(t)$ is determined  up to a factor
$h$, the physical scale corresponding to a given $s$ is
multiplied by $h^{-1}$, the same factors which enters all
measured  distances. Hence, with changing $h$, all our fits and
the data curve from  Figure 2 are to a good approximation simply
scaled horizontally by a  common factor.}.
The plots (a), (b) and (c) in Figure 3 correspond  to $p=0.11$,
$0.13$,  and $0.15$
respectively. They are all calculated using $\alpha =10$ and
$\gamma_s=10$, realistic values for these parameters as verified
by direct computations.
A selection of numerical values of $\frac{\delta
\rho}{\rho}$ for the three   different
curves in Figure 3 and the corresponding values for the curve in
Figure 2  are shown in Table 1.
The sets of parameters of the potential corresponding to the
different curves in Figure 3 are given in Table 2.
All the plots are normalized in such a
way that their peaks correspond to the scale of 30 Mpc (which
fixes the mass parameter $m$ for given $p$) and that their
maximal height is 0.187  (which in turn
fixes the height of the potential or the symmetry breaking scale
$\upsilon$  for fixed $p$, $m$ and $\alpha$). One can see that
the fits give very tiny masses in all the cases and symmetry
breaking scales $\upsilon$ of about $6  \times  10^{13} $ GeV.
It follows from (\ref{hi})
 that the associated values
of $H_i$ are of the order $ 10^{14}$ GeV,
 near the
present upper limits on the inflationary Hubble constant. It is
interesting to note that these values of parameters may arise
in a wide range of well motivated particle models.
Even such tiny masses as the ones discussed here may be stable
against radiative corrections if protected by exact or nearly
exact symmetries as happens in the models of ref [6],\cite{lr}.

Analyzing the plots (a), (b), (c) in Figure 3
one can  see that the smaller the probability $p$ is the larger the
slope of the curve at small and large scales. The character of
the curves may be understood by noting that they represent
the superposition of two  effects acting in opposite directions.
One is the number of s-clusters which is large at small scales
(hence, small s) and falls down
 quasi-exponentially toward
large scales. The other is the amount of energy stored in the
boundary of an s-cluster which is proportional to s, hence small
at small scales and  rising toward large ones. Table 2 contains
the values of $s$ which correspond to the lower and upper
limiting  scales considered in this paper.  First of all, one
can see that the s giving rise to perturbations at the lower end
of the domain of our plots, $s(20 \;{\rm Mpc})$, are in all the cases
larger than the  respective percolation correlation length
$s_{\xi}$. In fact, it is possible  to continue the percolation
spectra down below $s_{\xi}$, but this requires some further
analysis at the level of percolation theory, which is beyond the
scope of this work. For this reason we do not attempt here to compare
our results to QDOT data for scales below 20 Mpc.
Next, one can easily check that the values
of s  corresponding to
 the upper edge of
the scale interval we consider here, see the $s_{COBE}$ column
in Table 2, fall within the  domain of applicability of the
 percolation formulae listed in
Section 3.

All the curves in Figure 3 are normalized to fit the difference
between  QDOT data and CDM predictions on
scales smaller than 80 Mpc. It is clear that density
fluctuations induced by structure transitions accurately account
for this difference, both in magnitude and in the shape of the
distribution. We point out that this distribution is a natural
consequence of three-dimensional percolation theory, its shape
being set by the basic laws governing percolation.

 An important independent check for
our statistical picture of large scale structure formation is the
computation of density perturbation at the COBE scale (for h=1 it is
730 Mpc) and  comparison of these
numbers with the result reported by Smoot et al. [1],
$\frac{\delta T}{T} |_{COBE} = 1.1  \times  10^{-5} $.
The volume $s_{COBE}$ of clusters responsible for fluctuations
at the scale
of 730 Mpc is found easily by inverting the formula (\ref{dist})
and using (\ref{hzt}), (\ref{zas}), and (\ref{ztt}). If one computes
$z_a (s_{COBE})$, then it turns out that this redshift happens to be
much larger than $z_d$. This means that clusters producing fluctuations
at the COBE scale disappear long before the matter dominance epoch.
It follows that gravitational potential fluctuations which give
the dominant contribution to the microwave background anisotropy
at large scales
come from fluctuations induced by walls in the surrounding matter,
both dark
and baryonic, and not from actual domain walls entering the horizon
at the time of
decoupling or later.
It may easily be demonstrated that the direct contribution to
fluctuations given by walls whose radii become smaller
than the FRW horizon at
small redshifts is subdominant by computing the number of such walls
expected within the visible horizon. This is done using formula (\ref{zas}),
assuming $z_a (s)$ to be of the order of $z_d$ or smaller and
then using (\ref{ns})
and (\ref{nn}). The numbers corresponding to $z_a =z_d$ are given
in Table 3 in the columns labelled $s_a$, the volume of the cluster whose
radius of gyration enters horizon at $z_d$, and $N_{s_a}$, the total
number of such clusters expected. It is clear that
the contribution to the microwave
background anisotropy at large scales coming from large bubbles is
very strongly statistically suppressed (for further discussion of this
point see section 7)
and subdominant with respect to the effect
resulting from wall-induced fluctuations in dark and baryonic
matter.

As may be seen from Table 2, all our
cases can easily account for the COBE measurements, predicting
fluctuations  of $\delta T /T$ exactly of the required order of
magnitude (we use the  standard
Sachs-Wolfe formula to convert
 $\delta \rho / \rho$ into $\delta T /T$)\footnote[3]{Of course,
as discussed  for instance in [13], [14], given a model of
inflation one can try to explain the COBE
measurement by gravitational waves produced during inflation, but
the  magnitude of this effect is strongly model dependent. Since
we do not rely in our analysis on any specific model of
inflation, there is no inconsistency  in obtaining, at the COBE
scale, domain wall contributions  comparable to the COBE
result. We can always restrict ourselves to models
with a low power polynomial potential, like chaotic inflation,
where the gravitational waves contribution
can be shown to be subdominant [14].}.
In addition to the magnitude of the
$\delta T / T$ anisotropy, the COBE experiment [1] also gives the value
of the fluctuation spectral index n=$1.15 \pm 0.6\;$. Index n is defined
through the power-law ansatz for the power spectrum of primordial
fluctuations $P(k)=|\delta_{k}|^2 = A k^n $, where $\delta_{k}$ is the
Fourier transform of the overdensity field $\delta \rho / \rho$.
However, as analyzed in detail by Scaramella and Vittorio [16],
to judge the viability
of models predicting specific values of n one has to take
carefully into account
the effect of cosmic variance. The reason for this is that different
cosmic observers look at different skies which, in turn, are different
realizations
of the density field. One has to ask the question of how many observers
would see a realization of the sky consistent with data for a value
of n predicted by a given model. It is shown in reference [16] that
more than 50\% of the observers could see a sky which gives a
good fit to the COBE data
as long as $-1.5 < n < 2.5$.
In the present case, given the analytical formula (\ref{sm}), we can easily
compute the best values of n for different percolation probabilities $p$.
The values which give the best fits of the power law ansatz for the
power spectrum to our model between the COBE scale, 730 Mpc, and 2000 Mpc
are listed in the last column of the Table 2. Obviously, all the
cases we consider lie well within the limits given by Scaramella and
Vittorio. We note that the larger the probability $p$ the closer is the
predicted spectral
index n to the value n=$1.15$.

It should be stressed that our results are obtained without any
extreme fine-tuning. Curves corresponding to $p$ between 0.11 and
0.15 are very close to each other and similar stability of the
results holds with respect to small changes in all remaining
parameters. For example, in all the above cases we  have
taken $\gamma_s = 10$. One should note, however, that even taking
$\gamma_s = 1$ gives reasonable model parameters and
 it is clear that taking $\gamma_s
> 10$ would produce acceptable examples of models with smaller
values of the  symmetry breaking scale $\upsilon$ required to fit
the magnitude of the  experimental spectrum.
\\
\\
\noindent{\rm \large 7. Microwave background distorsions}
\vspace{2ex}

The walls that disappear by $z_d$ (well before decoupling of
photons) do not cause any direct perturbation to the microwave
background. They induce energy density perturbations in
background matter (dark and baryonic) and those perturbations are
standard effects, believed to be innocuous if the resulting
amplitude of theoretical fluctuations agrees with the observed
one  (it is our attitude to normalize our spectra in exactly this
way).  There are, however, additional effects due to walls which
annihilate late, during the epoch of recombination and after.
Radii of these bubbles  are of the order of the horizon scale at
decoupling and larger  ($\geq 100$ Mpc at
present), and, as discussed in [15], such bubbles collapsing
after decoupling form a time dependent gravitational potential
which produces a shift in the frequencies of the photons passing through
them. The estimate of corresponding distorsions to the $\delta
T /T$ of the photon background, assuming that walls are actually
present within our horizon, is [15] \beq
\frac{\delta T}{T} \approx 8 \pi G \sigma H_{0}^{-1}
\label{lim}
\eeq
Demanding that $\delta T /T \leq 5  \times 10^{-5}$ (the limits
on $\delta T / T$ at  small angular scales $\theta < 1^{o}$ are
less restrictive than at large scales) and naively using
(\ref{lim}) one gets
\beq
\sigma \leq 5  \times  10^{-9} {\rm GeV}^3
\label{mev}
\eeq
which is rather restrictive (see Table 3) and at first sight
cannot be satisfied
in the  examples discussed here. However, our calculation tells
us that the probability of seeing such a large bag  of wall is
extremely  tiny. In fact, the actual $\delta T / T$ due to these
walls  should be weighted by the square root of the average
number of a given size wall-bags on our lattice, $\sqrt{N_s}$,
assuming that photons perform a random walk among wall-bags. This
weighting number is always smaller than $\sqrt{N_{s_{a}}}$, where
$s_{a}$ is the volume of the cluster which enters  the horizon
at $z_d$\footnote[4]{In this paper $z_d$ denotes the  beginning
of matter
dominance. However, since the last scattering surface has a significant
thickness, and since its redshift is not that much smaller
than $z_d$, we  compute all the limits here using $z_d$, which
is a rather conservative  procedure in the present case.}. If we
weight in this way the right hand side of  (\ref{lim}), we then get
numbers respecting the above quoted upper limit on  $\delta T /T$
with a wide margin in all of our examples, as may easily be seen
from Table 3.
\\
\\
\noindent{\rm \large 8. Conclusions}
\vspace{2ex}

In this paper we have described the contribution to cosmological
energy density
fluctuations coming from  topological defects in the form of
domain walls produced during a biased phase transition in the
vacuum  of a light,  weakly interacting scalar field,
which we call the structuron. We have used as  an example a
transition triggered by inflationary quantum fluctuations, but
the shape of spatial statistics of the defects, and consequently  the
spectrum of their contribution to the overall energy density
fluctuations, will be similar for any
other biased phase transition. Using percolation theory methods
we have been able to describe in an analytical way the average
properties of the spatial  pattern formed by walls after a
transition of this type. Applying the  results to a simple cosine
potential, naturally arising in particle models  with light
scalars, we are able to produce an additional contribution  to
the mean cosmological energy density which has sufficient power
at larger  scales to account for the apparent difference between
the cold dark matter  spectrum normalized at small scales and the
total IRAS spectrum. As a bonus, after
continuation of our spectra beyond the range of  small and
intermediate scales, which they successfully describe, up to the
COBE scale, we find at that scale a contribution which is
precisely of the  order of the COBE result. Thus, structure
transitions offer a nontrivial
way of understanding nonzero microwave background distorsions at
very large cosmological scales. Also note that the nonequilibrium
phase transitions which we discuss,  the structure
transitions, naturally occur well before decoupling, deep in
the radiation  dominated epoch, and, therefore, can evade
problems that arise in
models with truly late time  phase transitions.

It is worth of pointing out that using phase transitions which
allow a percolative description of the resulting vacuum
structure for generating large scale
cosmological structure is a nontrivial and  novel approach. In
standard scenarios to explain  physical phenomena at a given
scale, one needs some causal physics operating at that scale, or
in other words,  significant correlations extending over that
scale. In our approach we exclusively use small scale
correlations operating over a fundamental  domain whose size is
orders of magnitude smaller than the distance to the FRW horizon. The
whole large scale pattern arises as a random, stochastic
superposition of independent small scale choices. It is
the astonishing property of such a statistical system that it
displays regular features, described by percolation theory,
which closely  correspond to regularities observed in the
distribution of matter in the  Universe.\\
\\
\\
\noindent{\large Acknowledgements}
\vspace{2ex}

Z. Lalak was supported by The Royal Society and by Alexander von
Humboldt Foundation. S. Lola was supported by the Greek State Scholarships
Foundation.\\
\\
\\

\newpage

\begin{center}

\begin{tabular}{ccccc}            \hline \hline
 X $h^{-1}$Mpc & {} & Y & {} & {}\\
 {} &$ \Delta$ & p=0.11 & p=0.13 & p=0.15 \\ \hline
 20 & 0.18 & 0.178  & 0.181 & 0.173 \\
 30 & 0.181& 0.187& 0.187& 0.187 \\
 40 & 0.174& 0.170& 0.169& 0.174 \\
 50 & 0.149& 0.142& 0.136& 0.149 \\
 63 & 0.117& 0.103& 0.1& 0.109 \\
 80 & 0.09 & 0.067& 0.07& 0.073 \\
 \hline
 \hline
\end{tabular}

\end{center}
\begin{center}
Table 1
\end{center}
\vspace{1cm}
\begin{center}

\begin{tabular}{cccccccc}            \hline \hline
$p$   & $m$ GeV   & $\upsilon$ GeV & $s_{\xi}$   &
$s(20\;Mpc)$ & $s_{COBE}$ & $\frac{\delta T}{T}|_{COBE}$ & $n$\\ \hline
0.11 &$3 \times 10^{-32}$  &$6.3 \times 10^{13}$  &$2.48$
&$3.3$&$39.6$&$1.4 \times 10^{-5}$&$2.20$ \\ 0.13 &$5 \times 10^{-32}$
&$6.3 \times 10^{13}$  &$3.08$ &$4.5$&$50.5$&$1.3 \times 10^{-5}$&$2.12$
\\ 0.15 &$4.5 \times 10^{-32}$&$7.1 \times 10^{13}$
&$3.93$&$4.36$&$59.8$& $1.7 \times 10^{-5}$&$2.06$\\ \hline
 \hline
\end{tabular}

\end{center}
\begin{center}
Table 2
\end{center}
\vspace{1cm}
\begin{center}

\begin{tabular}{cccc}            \hline \hline
$p$   & $s_a$   & $N_{s_{a}} $  & $\sigma \;{\rm GeV}^3$  \\
\hline 0.11 &$843$  &$8 \times 10^{-65}$  &$2.4 \times 10^{-4}$
\\ 0.13 &$1385$  &$2 \times 10^{-66}$  &$0.4 \times 10^{-3}$ \\
0.15 &$1385$&$5 \times 10^{-66}$  &$0.45 \times 10^{-3}$ \\
\hline \hline \end{tabular}

\end{center}
\begin{center}
Table 3
\end{center}

\newpage
\noindent{\large Figure Captions}\\
{}\\
{}\\
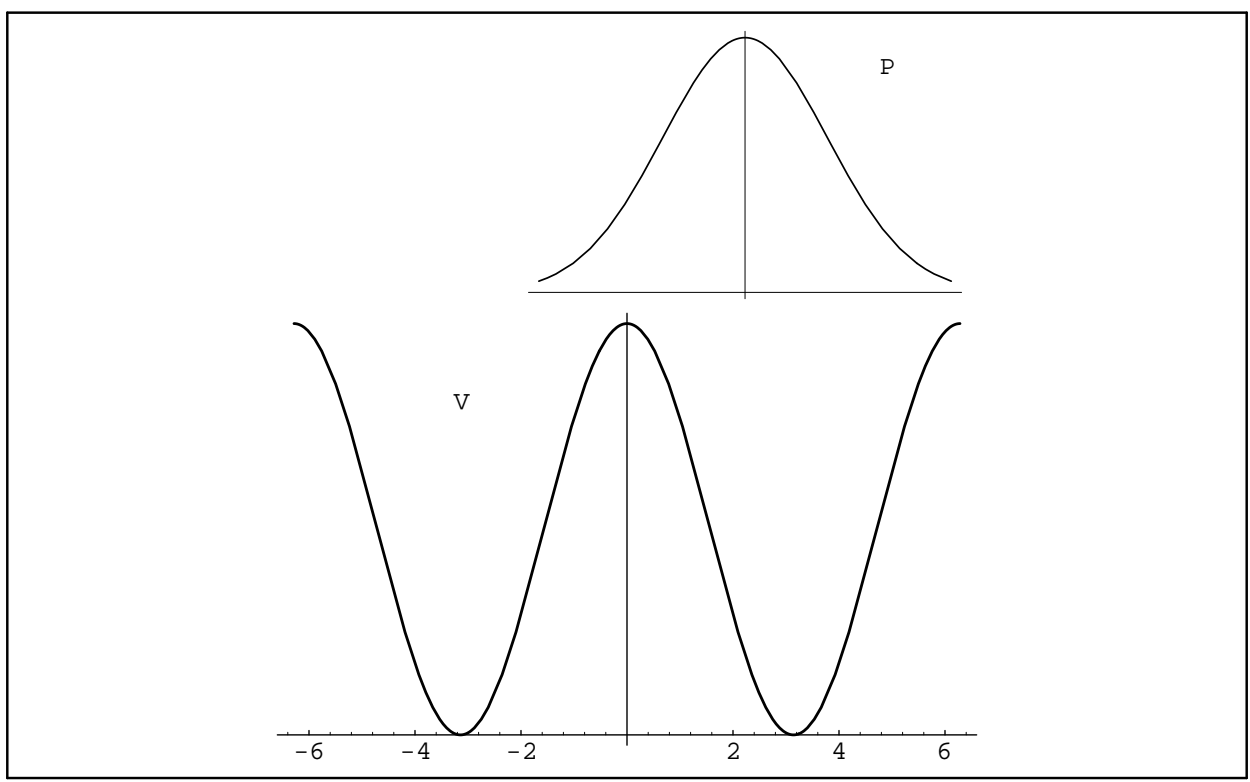
Figure 1. The probability distribution P of the random field
$\phi_q$ around an arbitrary mean value $\vartheta$ superimposed
on top of the  cosine potential
V. The horizontal axis gives values of $\phi / \vartheta$, the
units on the  vertical axes are arbitrary. Two minima of the
potential are shown, the left minimum corresponds to the (+)
vacuum, the right one to the (--) vacuum. \\
{}\\
{}\\
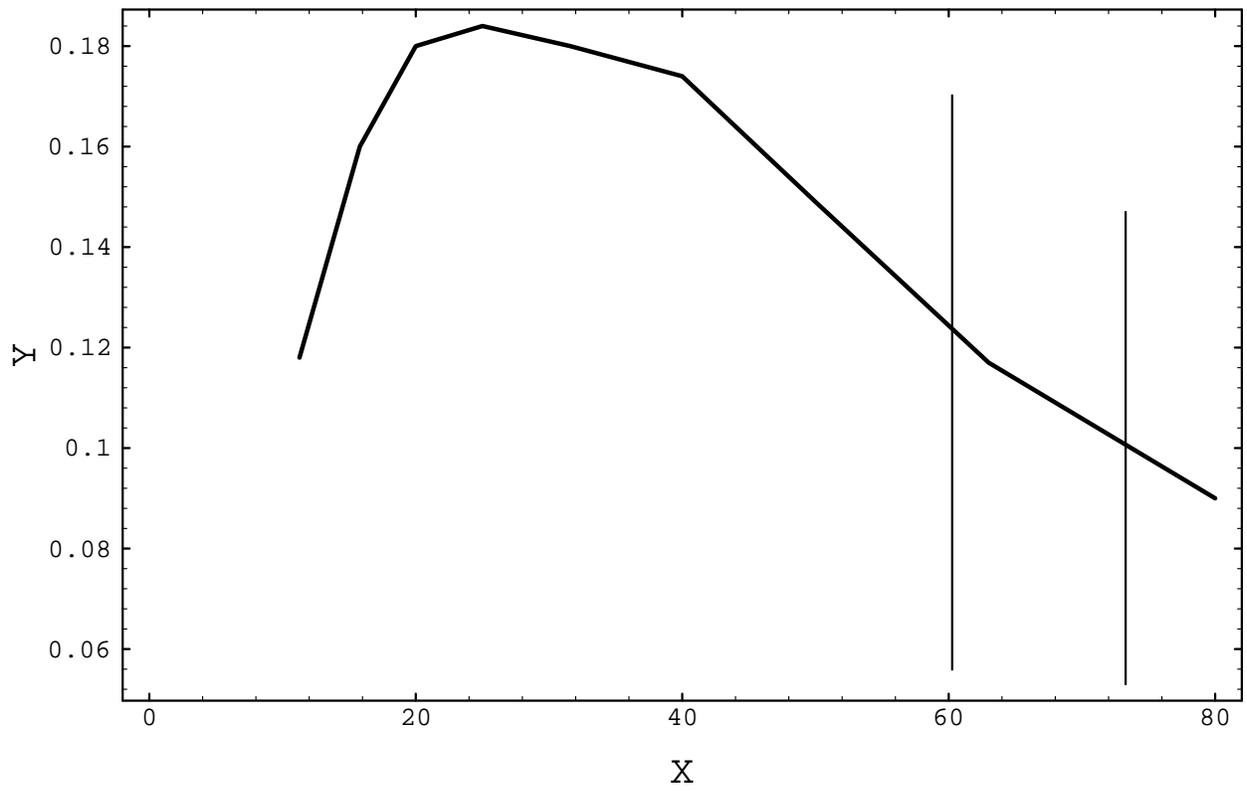
Figure 2. The difference between energy density overdensities,
$\left. \frac{\delta \rho}{\rho} \right |_{QDOT} -
\left. \frac{\delta \rho}{\rho} \right |_{CDM}$,
given by the QDOT analysis of the IRAS data and generated in
the CDM
scenario with a Harrison-Zeldovich spectrum of primeval
fluctuations [4]. Axis X represents physical scales in
Mpc$\;h^{-1}$, axis Y gives values of the difference between
overdensities. The two vertical lines represent uncertainties of
measurements at larger scales.\\
{}\\
{}\\
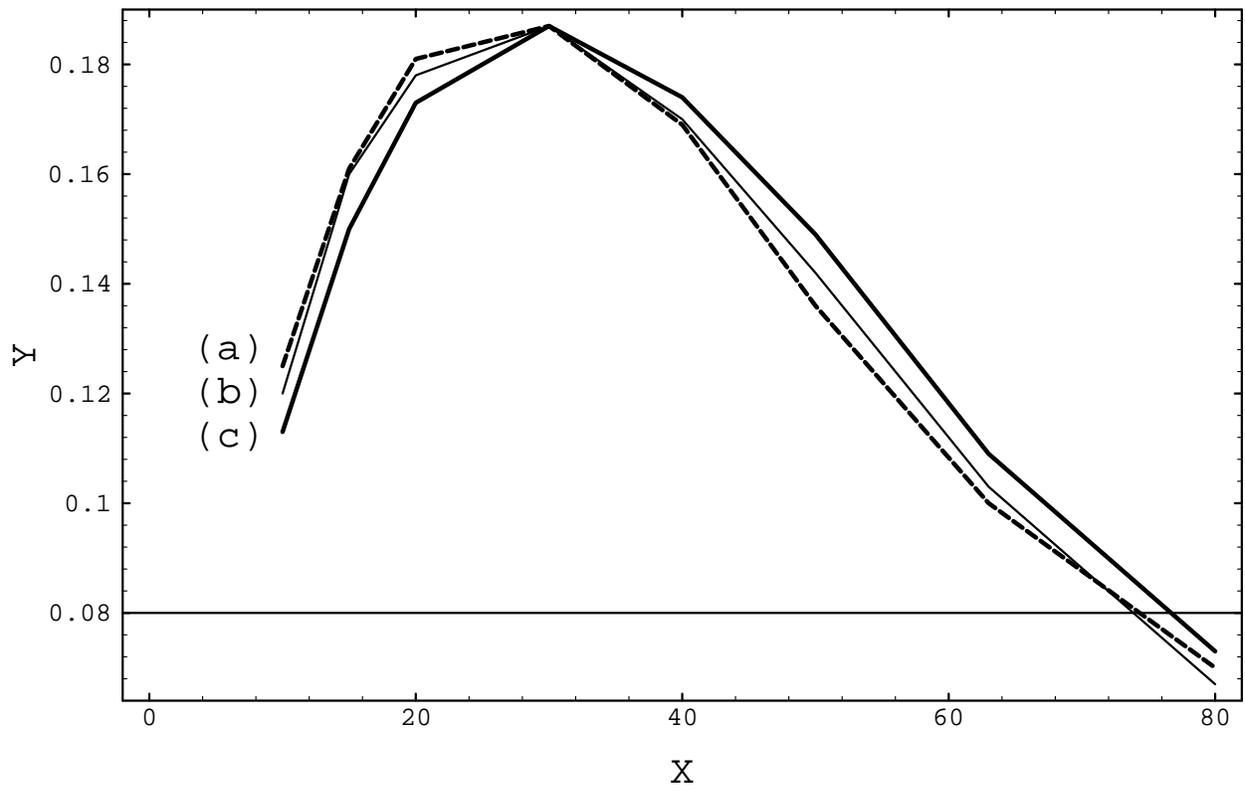
Figure 3. Wall-induced energy overdensity spectra in models
defined in Tables 2 and 3. Curves (a), (b), (c) correspond to
$p$=0.11, 0.13, 0.15 respectively. Axis X represents physical
scales in Mpc$\;h^{-1}$, axis Y gives values of overdensities.\\
{}\\
{}\\
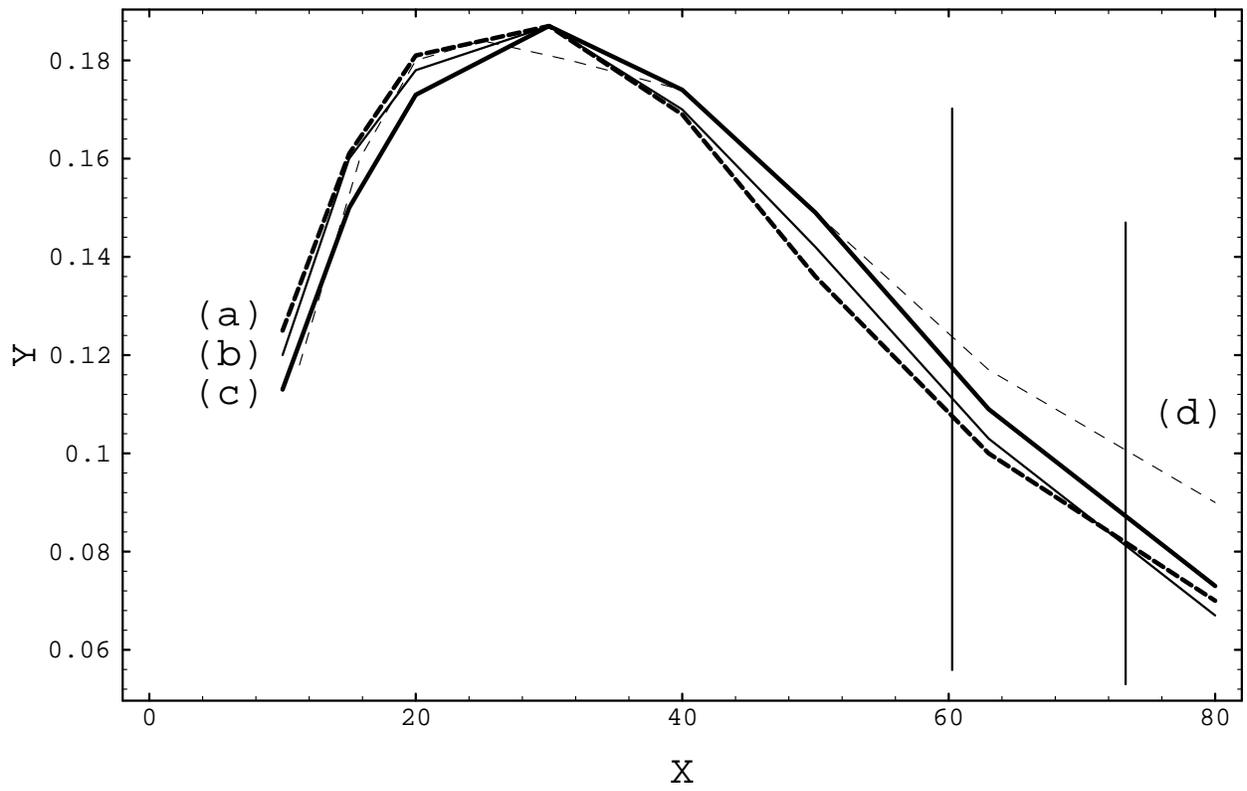
Figure 4. Superposition of plots from Figures 2 and 3. The dashed
curve  (d) is the QDOT data curve from Figure 3. \\
{}\\
{}\\
\newpage
\noindent{\large Table Captions}\\
{}\\
{}\\
Table 1. Numerical values of overdensities depicted in the
Figures 2  and 3. The column labelled $\Delta$ corresponds to the
curve from the  Figure 2 and contains data taken from [4].\\
{}\\
{}\\
Table 2. Parameters of the models generating spectra from the
Figure 3. The last two columns give predictions for the magnitude of the
microwave background anisotropy  at the COBE scale and for the
spectral index of fluctuations at large scales.\\
{}\\
{}\\
Table 3. This table illustrates statistical suppression of population
of large  bubbles in the percolation theory picture. The last column
gives values of the wall tension in models discussed in the
paper.\\ {}\\
{}\\

\end{document}